\documentclass{pasj00}

\title{Cosmic rays acceleration at ultrarelativistic subshocks}
\author{Janusz Bednarz}
\affil{Uniwersytet Warmi\'nsko-Mazurski, Wydzia\l{} Matematyki i Informatyki,
ul. \.Zo\l{}nierska 14, 10-561 Olsztyn, Poland}
\email{bednarz@matman.uwm.edu.pl}

\begin{document}

\maketitle

\begin{abstract}
We present the model of cosmic rays acceleration at ultrarelativistic subshocks
and confront it with the observations of gamma-ray bursts (GRBs) and blazars.
We investigate cosmic rays acceleration in shocks with Lorentz factors ($\gamma$)
in the range 3 - 40. We show that fluctuations of the magnetic field downstream
of the shock do not play an important role in the acceleration process. Results
of numerical simulations for shocks with considered Lorentz factors and
perpendicular magnetic field inclinations are presented. We fit the derived
particle energy spectral index ($\sigma$) dependence on fluctuations
of the magnetic field upstream and $\gamma$ with a function.
\end{abstract}

\section{Introduction}
The acceleration mechanism which operate at ultrarelativistic shock fronts
was discovered by Bednarz \& Ostrowski (1998). The mechanism is
different from the diffusive particle acceleration which is assumed
to be suppressed at superluminal shock fronts (Bell 1978; Drury 1983). 
The effect of the magnetic field direction is important for ultrarelativistic
shocks because all of them are superluminal. Medvedev \& Loeb (1999) have shown
that the relativistic two-stream instability naturally generate strong magnetic
fields which are parallel to the shock front. Therefore ultrarelativistic
shocks have to be superluminal even if one could imagine
an external magnetic field with the angle between the upstream field and the
shock normal smaller than $\sim 1/\gamma$. Ultrarelativistic shocks without
any mean magnetic fields cannot be considered as real physical phenomena by
the same reason.

It have appeared a few papers about a particle acceleration
at ultrarelativistic shocks without mean magnetic fields (Kirk et al. 2000;
Achterberg et al. 2001; Vietri 2003; Lemoine \& Pelletier 2003)
or with subluminal shock geometry (Ellison \& Double 2002 - parallel shocks).
Their acceleration is similar to the diffusive shock acceleration but it includes
anisotropies in the angular distribution upstream of the shock. However,
particles in this acceleration are able to return to the shock from downstream to
upstream due to large magnetic field fluctuations downstream of the shock or due
to subluminal shock geometry as in non-relativistic and mildly relativistic regime.
Thus, they have failed to understand the actual ultrarelativistic shock
acceleration mechanism because the returning is due to
small fluctuations of the magnetic field upstream of the shock (the needed
fluctuations decrease when the Lorentz factor of the shock increases)
and relativistic effects providing a small change of the particle trajectory
in the mean field upstream of the shock
to be large as measured downstream.
To date, the only numerical calculations performed by Bednarz \& Ostrowski (1998), 
Bednarz (2000) and in this paper consider the problem of particle acceleration
in ultrarelativistic shocks.

It is known for a long time that relativistic shocks occur in regions of
efficient acceleration of leptons. The acceleration to non-thermal distributions
is observed at hot spots of extragalactic radio sources, in blazars, GRBs
and pulsar wind nebulae. In order to account for the presence of these high
energy leptons, some authors have tried to find the acceleration mechanism.
Begelman \& Kirk (1990) proposed shock-drift acceleration at relativistic
shocks to operate at hot spots of extragalactic radio sources.
In the mechanism, particles are accelerated in a single shock crossing by
drifting parallel (or anti-parallel) to the electric field. Afterwards, Hoshino
et al. (1992) described a process of shock acceleration of positrons to non-thermal
distributions devoted to account for the synchrotron radiation observed in the Crab Nebula
and hot spots. In the model, the gyrating reflected protons downstream of the shock
dissipate their energy in the form of collectively emitted, left-handed magnetosonic
waves which are resonantly absorbed by the positrons immediately behind the ion
reflection region.
The dynamics of the Weibel instability has recently been simulated by several
research groups using 3D plasma kinetic code. These simulations confirm both the
generation of the magnetic field and the particle acceleration downstream of the shock.
The particle energising in Silva et al. (2003) simulations (electron-positron plasma)
is due to pitch angle scattering in the produced magnetic field after saturation.
The energy stored in the magnetic field is transfered back to the plasma particles.
Simulations of Frederiksen et al. (2003) have yielded the energy transfer from
protons to leptons similar to Hoshino et al. (1992). In their description,
the scattered protons create a fluctuating electric field which tends to
equilibrate the energy between protons and electrons. Nishikawa et al. (2003)
results suggest that electrons and ions are accelerated
in the direction transverse to the shock normal only.

All the described mechanisms suffer from small energies the particles are
able to acquire. The shock-drift acceleration allow for the energy increase of
about ten times. Leptons can receive the energy from protons which is
about ten times ($\sim\gamma$ times in Frederiksen et al. 2003 does not necessarily
depend on $\gamma$) above the thermal energy downstream of the shock.
Silva et al. (2003) simulations have led to the generation of high-energy tails
in the distribution function, with energies few times above
the thermal energy downstream of the shock. In our model presented below, we
apply some of these mechanisms to production of seed particles.

\section{The model of cosmic rays acceleration at ultrarelativistic subshocks}
In the model, we assume an extremely relativistic outflow
of electrons and positrons and probably protons and nuclei which
originates from a central engine. The outflow consists of few portions
of uniform plasma. These portions have different Lorentz factors and
relative Lorentz factors between some of them reach value from several up
to tens. Temperatures of these portions are similar to each other.

In the first phase of the acceleration process, leptons and baryonic
matter are preaccelerating either in the shock generated by the first slab
of the outflow hitting the material around the central object or in the
shock created by two colliding slabs. Depending on $\gamma$, seed particles
are producing in the forward and reverse shock or in the forward shock only.

After the collision the temperature of the resulting plasma is higher than
temperatures of the two slabs before the collision. We expect that the
external magnetic field influences to the dynamics of the Weibel instability
in such a way that the magnetic field produced by the instability is
oriented in the form of a toroidal magnetic field.
It has been suggested that hoop stresses associated with toroidal
fields wrapped around the extragalactic jet will pinch the outflowing
plasma and exert in this way a collimating force on the jet. Following
that example, we expect that the heating plasma will only expand in the
direction parallel to its velocity. In result, the collision yields a very
long and hot jet filled with seed particles and toroidal magnetic field.

We distinguish between dense enough and very thin material
of electrons, protons and probably nuclei surrounding the site of
the central engine. If the material is very thin, then protons or nuclei
must be present in the outflow to allow for leptons acceleration thanks to the
energy transfer from baryons (mechanisms: Hoshino et al. 1992, Frederiksen et al. 2003).
If the preacceleration takes place in two colliding slabs of electron-positron
plasma, then Silva et al. (2003) mechanism is being the only one at work.

In the second phase, a following slab of plasma is catching up with the
plasma filled with non-thermal relativistic particles. These particles
are rapidly accelerating in the ultrarelativistic shock acceleration mechanism
if the Lorentz factor of the formed shock is large enough. Afterwards, they flow
downstream of the shock and continually radiate synchrotron radiation.
The maximum energy the particles are able to reach is determined by the size
of the shock or by the value of the magnetic field fluctuations upstream of the shock
and $\gamma$ or by the acceleration time (it is very short, see Bednarz 2000)
and the plasma rest frame magnetic and radiative energy density.
However, only a few per cent of seed particles are involved in the process
by reason of small efficiency of cosmic ray reflections
(Bednarz \& Ostrowski 1999).

The presented scenario is capable of explaining many features of GRBs
and some features of blazars as well. The occurrence of the two phases
of the acceleration is one of them. Ghisellini et al. (2002) have derived
from spectral properties of blazars that a phase of pre-heating and
a phase of rapid acceleration leading to a non-thermal distribution
occur by turns in these objects. In GRBs, the first phase is observed
in the form of the precursor which was detected in a few per cent
of GRBs only (Koshut et al. 1995). The rare occurrence of precursors
is best explained by small energy of seed particles that strongly depends
on $\gamma$.

In the main phase of the acceleration, the shock slows down in
consequence of collecting slower material from upstream.
In the phase we distinguish two sub-phases. First, both the reverse
and the forward shocks are present and the speed of the forward
shock is constant. Next, the reverse shock disappears on the edge
of the back slab and the forward shock slows down.
The long and hot slab of plasma filled with seed particles and toroidal
magnetic field is much longer than the slab of plasma originating
from the central engine. In consequence, the reverse shock disappears
in much shorter time than the forward shock crosses the plasma and we can
neglect the first sub-phase.

Further on, we
consider that the maximum energy of particles in GRBs is determined
by the value of magnetic field fluctuations upstream of the shock
and $\gamma$ (we neglect the fraction of particles with large spectral
indices) and consider constant spectrum of magnetic field fluctuations
upstream of the shock. As will be shown further, the spectral index
of accelerating particles with momenta larger than a threshold momentum $p_t$
increases and simultaneously, it remains for the others approximately
constant as long
as $\gamma$ decreases. Moreover, $p_t$ decreases when the shock slows down.
The rising phase of the GRB pulse is attributed to the energising
of seed particles but the peak in the light curve is shifted
in different energy channels. The GRB lag (Kocevski \& Liang 2003)
arises from decreasing $p_t$. When first particles reach
$p_t(t=0)$ with $\sigma \approx \sigma_l$ (one should choose $\sigma$
slightly larger than $\sigma_l$ - the limiting spectral index),
the following ones reach $p_t(t=0)$ with $\sigma>\sigma_l$ because
$p_t(t>0)<p_t(t=0)$. In result, the excess of particles at $p_t$ moves
towards smaller momenta producing the observed lag.
Similarly, the detected spectral evolution of main pulses (Crider et al.
1997; Ghirlanda et al. 2002) arises due to increasing the energy of
particles with constant $\sigma \approx \sigma_l$ in the rising phase and
increasing $\sigma$ of particles with momenta larger than $p_t$ in the
decaying phase.

Many shapes of main pulses resemble a FRED (fast rise, exponential decay).
The rising phase of pulses produced by shocks with larger initial $\gamma$
must yield steeper light curves than the phase produced by slower shocks.
The main reason of this is the reduction in the compression of the magnetic
field which, in turn, causes the increase of the acceleration time
(equal to $\sim r_g/c$ - Bednarz 2000, $r_g$ - is the particle gyroradius
and $c$ - the speed of light). The decaying phase is long because particles
are still accelerating but it can be fast if the shock producing the main
pulse comes to the region without seed particles before particles reach $p_t$.
The reverse shock could supply with seed particles less efficiently than the
forward shock can and the light curve will exhibit two peaks when a following
shock ploughs into such plasma. One could also imagine a break in the field of
seed particles if two shocks generate it independently. 

We propose that short variability timescale pulses, which overlap on main
pulses, arise from interaction between the magnetic field generated
by the shock and a strong external magnetic field or the toroidal field of
the hot forward plasma.
That interaction can produce fast changes in
direction of the shock velocity (similar to gyration) and, by turns,
the sub-pulses because the detected flux strongly depends on the angle
between the shock normal and the direction of observation.

Most of the particle spectral indices in blazars (Ghisellini et al. 2002) and
in GRBs are in good agreement with the value
of 2.18 and larger derived for ultrarelativistic shocks in this paper.
Suggesting smaller value of $\sigma$ (Panaitescu \& Kumar 2002) is
naturally explained by a jet model with continuous energy injection and
$\sigma$ above 2.18 (Bj\"ornsson et al. 2002). The continuous energy injection
is in agreement with the presented model. In some of the GRBs events, the last
slab (or a group) of plasma can be the slowest one. First, the distance between
the forward and the backward plasma is increasing and the forward plasma
is collecting interstellar matter and slowing down. Next, the distance is
decreasing and finally, the backward plasma hits the forward one and forms a shock.
We expect that, in many cases, $\gamma$ of the shock is large enough to allow for
the acceleration across the main mechanism. The forward material is in the form
of long, thin jet with the Lorentz factor considerably smaller than at the
GRB stage. The external magnetic field is much weaker than the field near the central
engine. All the reasons influence that the detecting radiation is below X-ray
range and is lasting for long time. The colliding jets are responsible for
breaks in the afterglow light curves.

There is an open question if relativistic shocks can produce seed
protons and nuclei. The simulations of Nishikawa et al. (2003) suggest that
they could.
Heavy particles in a cold plasma upstream of the shock are seeing a shock
generated by leptons as a discontinuity but they are not ultrarelativistic
there. We suggest that seed baryons could be produced in a mechanism similar
to shock-drift acceleration in these shocks if leptons taking away the energy
from baryons downstream of the shock remain the fraction of total baryons which
end up with super-thermal energies. In this picture, the ultrarelativistic shock
acceleration mechanism could produce a major part of the cosmic rays.

\section{Simulations}
In simulations we follow the procedure used by Bednarz \& Ostrowski (1996)
with a hybrid approach used by Bednarz \& Ostrowski (1998). Monoenergetic
seed particles are injected at the shock with the same initial weight factor.
Each particle trajectory is followed using numerical computations until the
particle
escapes through the free escape boundary placed far downstream from the shock
or it reaches the energy larger than the assumed upper limit. These particles
are replaced with the ones arising from splitting the remaining high-weight
particles in a way to preserve their phase space coordinates, but ascribing
only a half or a smaller respective part of the original particle weight to
each of the resulting particles. All computations are performed in the
respective upstream or downstream plasma rest frame. Each time the particle
crosses the shock its momentum is Lorentz transformed to the respective
plasma rest frame and, in the shock normal rest frame, the respective contribution
is added to the given momentum bin in the particle spectrum. We derive the particle
trajectories in the mean magnetic field and a fluctuating component. The perturbed
magnetic field represents the traditional picture based on the concept
of magnetic scattering centres. It is simulated by small
amplitude particle momentum scattering within a cone with angular opening
$\Delta \Omega$ less than the particle anisotropy $\sim 1/\gamma$.
The particle momentum scattering distribution is uniform
within the cone. The time (measured in $r_g/c$) between scatterings is uniformly
distributed from $0.5 t_{d,u}$ to $1.5 t_{d,u}$ ($t_d$ and $t_u$
are mean times between scatterings downstream and upstream respectively).
After crossing the shock, the remaining time the particle should follow
in the mean field before the next scattering is multiplied or divided by
$t_d/t_u$. Particle energy is preserved at each scattering in the plasma
rest frame. For the considered continuous injection after initial time, the energy
cut-off of the formed spectrum shifts toward higher energies with time.
The resulting spectra allow one to fit spectral indices and derive acceleration
time.
The downstream magnetic field, its inclination to the shock normal
and shock speed are derived for the relativistic shock with the
compression $R_h=3 U_1^2$ ($U_1$ is the shock speed) for the hot plasma.
In the simulations presented in Fig.~\ref{fig2}, we applied compression $R_c$
obtained with the formula of Heavens \& Drury (1988) for a
cold proton-electron plasma.

\begin{figure}
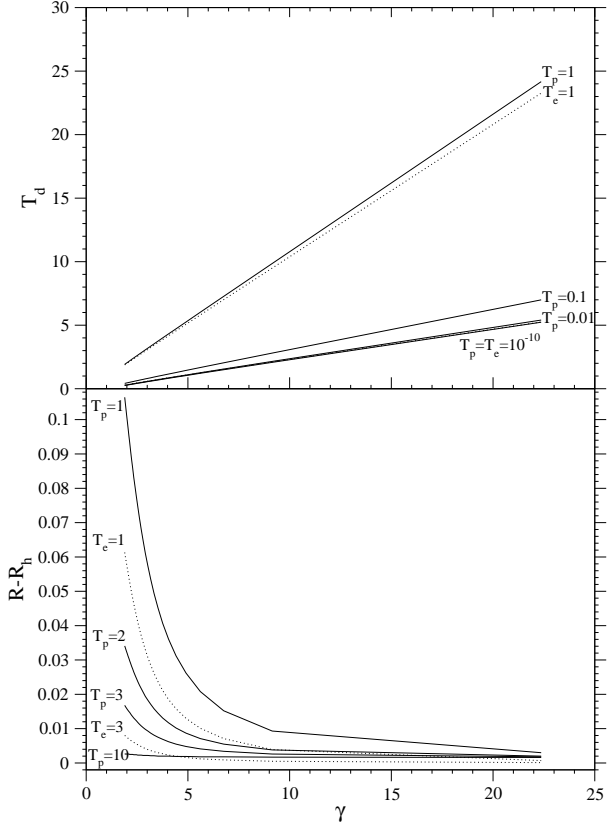

 \begin{center}
  \FigureFile(80mm,110mm){fig1.eps}
 \end{center}
\caption{The downstream temperature and the difference between
the compression and the compression of the hot plasma limit as a
function of the shock Lorentz factor.
The curves correspond to different upstream temperatures
and are marked with $T_p$ (solid curves) or $T_e$ (dotted curves).
In the upper panel, $T_p=T_e=10^{-10}$ is close to $T_p=0.01$.}
\label{fig1}
\end{figure}

In order to estimate the correctness of $R_h$ for slow shocks, we followed the
computations of Heavens \& Drury (1988). The bottom panel of Fig.~\ref{fig1} shows the
difference between $R$ (the real compression) and $R_h$
as a function of $\gamma$ for some cases of the upstream fluid ($T_p$, $T_e$
- the upstream temperature of proton-electron, electron-positron plasma in units of
$5.4\cdot 10^{12} K$, $5.9\cdot 10^{9} K$ respectively).
The top panel shows the relation between $\gamma$ and the downstream
temperature ($T_d$ - in the same unit as appropriate $T_{p,e}$). We can draw a conclusion
that $\gamma$ of the shock hitting a cold plasma must be larger than 6 during
the preacceleration so that $R_h$ might be adequate for shocks with $\gamma < 5$ in the main
phase.

\section{Some features of the acceleration mechanism}

\begin{figure}
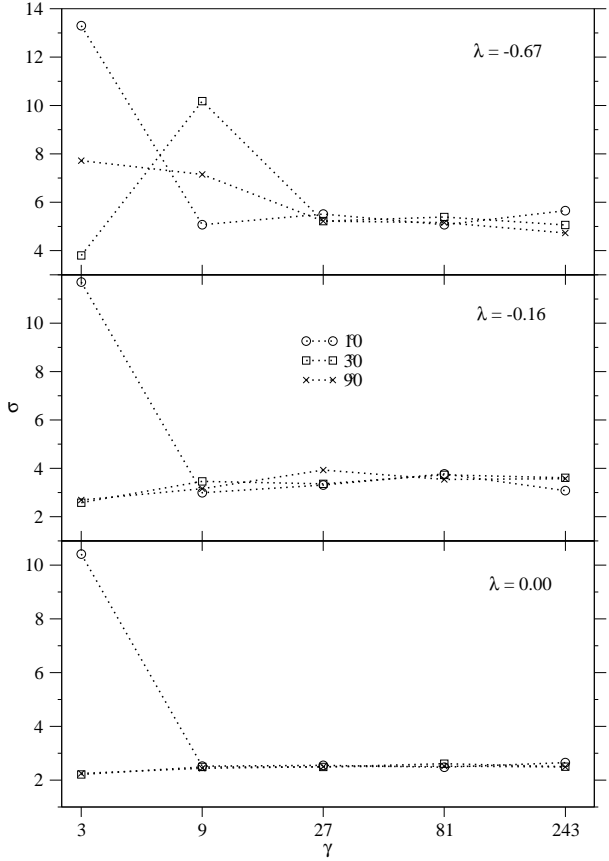

 \begin{center}
  \FigureFile(80mm,114mm){fig2.eps}
 \end{center}
\caption{Simulated spectral indices as a function of the shock Lorentz
factor. Three values of the magnetic field inclination upstream of
the shock are applied ($\psi = 10^\circ , 30^\circ , 90^\circ$). Magnetic field
fluctuations upstream are not present. Results for given
downstream magnetic field perturbations $\lambda$ are presented separately.}
\label{fig2}
\end{figure}

We have performed simulations in the absence of magnetic field fluctuations
upstream and for a few cases of the fluctuations
downstream measured by a parameter
$\lambda=log(\frac{\kappa_\perp}{\kappa_\parallel})$, where $\kappa_\perp$
and $\kappa_\parallel$ are perpendicular and parallel diffusion coefficients
respectively.
We have chosen a few values of $\gamma$ and the upstream magnetic field inclination
with respect to the shock normal $\psi = 10^\circ , 30^\circ , 90^\circ$.\
We exhibit the outcome of the simulations in Fig.~\ref{fig2} . The large value of $\sigma$
for $\psi = 10^\circ$ and $\gamma=3$ arises due to escaping a large fraction of
particles far upstream from the shock.
The $\sigma$ dependence on $\psi$ or $\gamma$ is present for large value of
$\sigma$ only.
We have obtained $\sigma=2.5; 3.5; 5$ for $\lambda=0; -0.16; -0.67$ respectively.

As was shown above, medium-value magnetic field fluctuations downstream
can be neglected in the acceleration. We suspect that they are really small
if we take into account very high linear polarization reported in RHESSI
observations of GRB021206 (Coburn \& Boggs 2003).
Moreover, the particles do not interact resonantly with magnetic field
perturbations upstream since the time between shock crossings downstream-upstream
and upstream-downstream is a small part of $r_g/c$.
Therefore, our scattering model is the best because it reproduces the non-resonant
interactions excellently and independently of magnetic field fluctuations spectra.
The model follows the relation
$\frac{\kappa_\perp}{\kappa_\parallel}=\frac{1}{1+const(\omega\tau)^2}$,
$\tau$ is the mean time when the particle momentum ($p$) direction shifts at
an angle smaller than $20^\circ$ because of scatterings
and $\omega$ is the gyration frequency.
Let us introduce the ordinary assumption that each fluctuation shifts the
direction at a small angle $\sim 1/p$. As a result, $\tau \sim p^2$ and
subsequently, $\frac{\kappa_\perp}{\kappa_\parallel}=\frac{1}{1+Qp^2}$
($Q$ results from fluctuations spectra).
For large $p$ we get $\lambda=-2log_{10}(Qp)$. Eq. (\ref{A}) yields that
$\lambda$ upstream increases ($p$ decreases) with decreasing $\gamma$ and
constant $\sigma$. That is the reason for the lag in GRBs light curves.

\begin{figure}
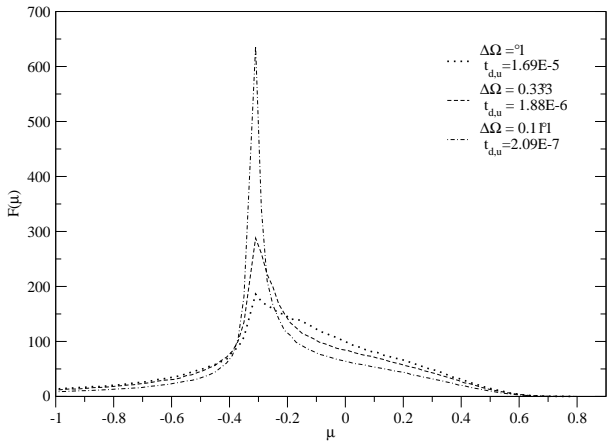

 \begin{center}
  \FigureFile(80mm,58mm){fig3.eps}
 \end{center}
\caption{Simulated particle angular distributions at the shock,
as measured in the downstream plasma rest frame, for $\gamma=27$,
$\psi=90^\circ$ and $\lambda=0$ (downstream and upstream). The
distribution function narrows when $t_{d,u}$ decreases but
$\sigma$ is constant.}
\label{fig3}
\end{figure}

The angular distribution of particles at the shock in the downstream plasma
rest frame presented by Bednarz \& Ostrowski (1998) peaks at $\mu = -U_2$,
$U_2$ is the shock speed in the downstream rest frame and $\mu$ is the
cosines of the angle between the shock normal and the particle momentum.
The distribution presented by other authors peaks at $\mu\simeq 0$ but
they did not follow full 3D simulations (or calculations) with the mean field
and we cannot compare the results because it is difficult for us to mimic
the reduced situations.
We expect the peak at $-U_2$ arises from keeping a fraction of
particles close to the shock. The particles contribute to the distribution
considerably because their velocities perpendicular to the shock plane approach
the shock speed.
They can oscillate between upstream and downstream especially, if scatterings
upstream are close to the shock plane.
In Fig.~\ref{fig3}, we present the simulations for $\gamma = 27$, $\psi = 90^\circ$,
different $t_{d,u}$ (the diffusion is the same because
$\Delta\Omega^2/t_{d,e}=const$) and $\lambda=0$.
The distribution function narrows with decreasing $t_{d,u}$ but $\sigma$
does not change.

The explanation of the shift in the distribution function presented by
Ostrowski \& Bednarz (2002) is false\footnote{That is not mine. M. Ostrowski did
not consult with me the context of the paper.}. It is obvious that the absolute
value of the inverse of the relative particle velocity with respect to the shock
is the particle weight. One obtains unlike distribution if one applies the velocity in
the plasma rest frame.

We obtained the limiting spectral index for $\lambda=0$ upstream and
downstream, $\psi=90^\circ$ and $\gamma$ from 60 to 120 (33 points).
It is equal to $\sigma_{lm}(\gamma>60) = 2.23 \pm 0.03$.


\section{The relation between $\sigma$, $\gamma$ and $\lambda$}

\begin{figure}
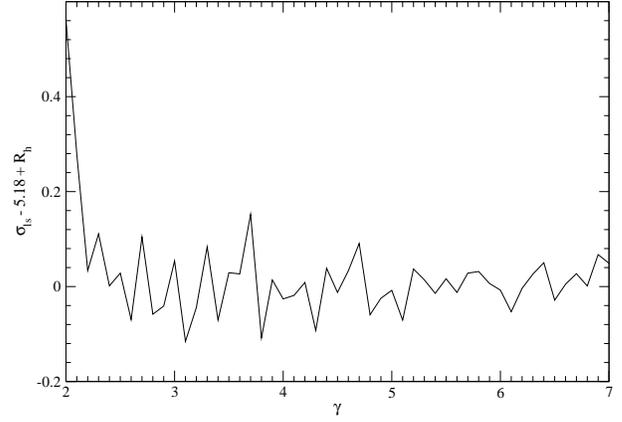

 \begin{center}
  \FigureFile(80mm,55mm){fig4.eps}
 \end{center}
\caption{ The difference between the limiting spectral index
obtained from simulations ($\sigma_{ls}$) and $\sigma_l$.}
\label{fig4}
\end{figure}

In the simulations below, we have applied $t_d =0.1$ and $\Delta \Omega = 10^{-6}$
downstream ($e^\lambda=0$, homogeneous magnetic field) and $\psi = 90^\circ$.
First, we have found the limiting spectral index for large $\gamma$ ($\gamma$ from
60 to 100 - 58 points, $\lambda=0$ upstream): $\sigma_{l0}(\gamma>60) = 2.18 \pm 0.03$.
Next, for the same conditions ($\psi = 90^\circ$ and $e^\lambda=0$ downstream),
our simulations have covered two $\gamma$ ranges: 1) from 2 to 7 every 0.1; 2) from 2 to 40
every 1 and $\lambda$ upstream from 0 to -4.3. Then, we have been able to obtain
$\sigma_l$ for $\gamma>3$: $\sigma_l = 5.18-R_h$ (see Fig.~\ref{fig4}).
We have continued analysis of the data. Let us introduce two
parameters $r_m=R_h/3$ and $r_d=R_h-2$.
We have fitted $\sigma/\sigma_l -r_m^{3.2}$ as a function of
$\lambda_p=3.36-1.94 log_{10}(\gamma)-0.3 r_d-\lambda/r_d$
for $\sigma<7$, $\lambda_p>0.55$ and $3\le \gamma \le 40$:

\begin{equation}
\begin{array}[c]{ll}
\frac{\sigma}{\sigma_l}-r_m^{3.2} = K (\lambda_p^5+13.1 \lambda_p )
(1-\frac{0.147}{\lambda_p}),\\
K = 4.6 \cdot 10^{-3} \pm \frac{2 \cdot 10^{-3}}{\lambda_p}.
\end{array}
\label{A}
\end{equation}

\noindent
The fitting is correct if $\lambda_p>0$ but the error is too large and
there are systematic effects if $\sigma \to \sigma_l$.
We have replaced (\ref{A}) with the linear function from
$\sigma \approx 1.03 \sigma_e$ ($\sigma_e$ is the limiting $\sigma$
derived from (\ref{A}))
to $\sigma_l$ to improve eq. (\ref{A}).
We have fixed the intersection of the two curves at
$\lambda_{pc}=\lambda_p(\sigma\approx 1.03 \sigma_e)$:

\begin{equation}
\begin{array}[c]{ll}
\lambda_{pc}= 2.86-0.265\gamma+0.015\gamma^2-4.02\cdot 10^{-4}\gamma^{3}+\\
+3.88\cdot 10^{-6}\gamma^4.
\end{array}
\label{B}
\end{equation}

\begin{figure}
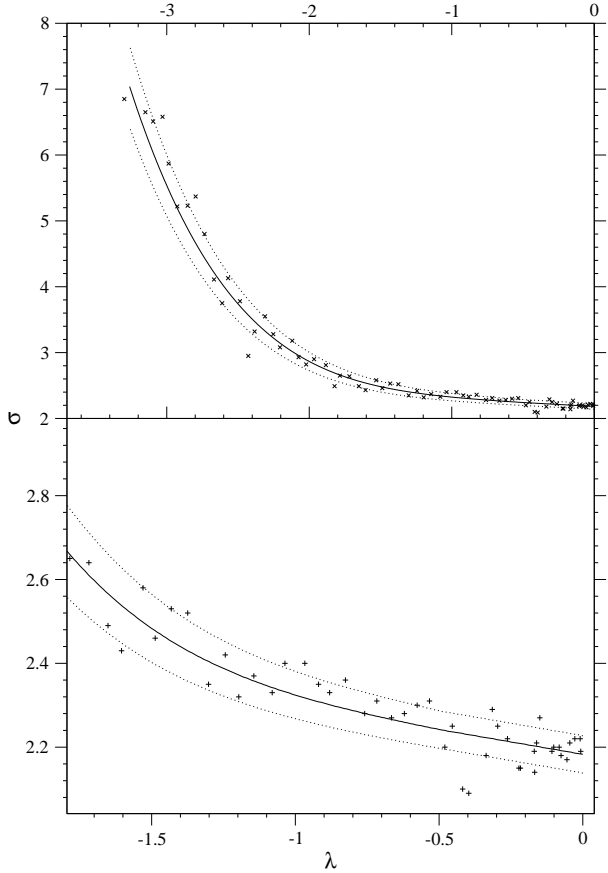

 \begin{center}
  \FigureFile(80mm,116mm){fig5.eps}
 \end{center}
\caption{ The spectral index of the energy distribution function at the shock,
for particles accelerated in the shock with $\gamma=32$ propagating into hot
plasma, as a function of magnetic field fluctuations upstream
($e^\lambda = 0$ downstream and $\psi = 90^\circ$).
The solid line is composed of two curves (equations (\ref{A}) and (\ref{C}))
with the intersection at (\ref{B}).
Dotted lines define the standard error and crosses represent results of
simulations.}
\label{fig5}
\end{figure}

\begin{figure}
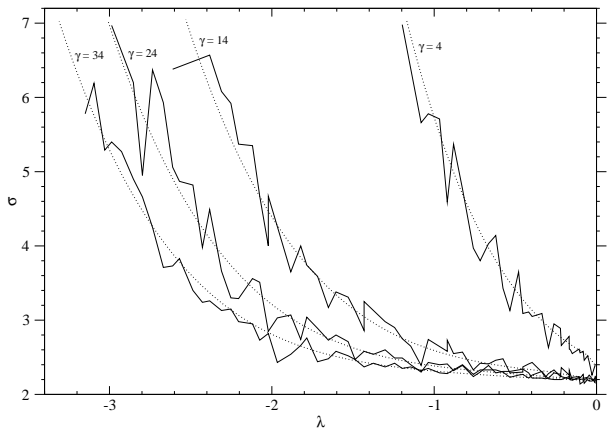

 \begin{center}
  \FigureFile(80mm,56mm){fig6.eps}
 \end{center}
\caption{ The spectral index of the energy distribution function at the shock,
for particles accelerated in the shock with $\gamma=4; 14; 24; 34$ propagating
into hot plasma, as a function of magnetic field fluctuations upstream
($e^\lambda = 0$ downstream and $\psi = 90^\circ$).
The numerical results are joined by the solid lines. The dotted lines
represent the fit.}
\label{fig6}
\end{figure}

\noindent
The linear function is

\begin{equation}
\begin{array}{l}
\sigma=\sigma_l(a\lambda+1)\pm \Delta\sigma ,\;\;\;\;\;
a=((\frac{\sigma}{\sigma_l})_c-1)/\lambda_c ,
\end{array}
\label{C}
\end{equation}

\noindent
where the select error
$\Delta\sigma=\sigma_l ((\frac{\sigma}{\sigma_l})_c-r_m^{3.2})/
(2.3 \lambda_{pc})$ depends on $\gamma$ only at the intersection
$(\frac{\sigma}{\sigma_l})_c=4.6\cdot 10^{-3}(\lambda_{pc}^5+13.1 \lambda_{pc} )
(1-\frac{0.147}{\lambda_{pc}})+r_m^{3.2}$, 
$\lambda_c=r_d(3.36-1.94log_{10}(\gamma)-0.3r_d-\lambda_{pc})$ and below.
An example of a curve derived from (\ref{A}), (\ref{B}) and (\ref{C}) for
$\gamma=32$ is presented in Fig.~\ref{fig5} and a few curves for different
$\gamma$ are shown in Fig.~\ref{fig6}. The relation between $\sigma$, $\gamma$
and $\lambda$ is necessary to model GRBs light curves.


\section*{Acknowledgements}

We acknowledge support from the {\it Komitet
Bada\'n Naukowych\/} through the grant 5P03D.007.20~.

\end{document}